\documentclass[useAMS,usenatbib]{mn2e}
\usepackage[utf8x]{inputenc}
\usepackage{hyperref}
\usepackage{amsmath}
\usepackage{graphicx}
\usepackage{booktabs,ctable}
\usepackage{subfigure}
\usepackage{color}

\def\be{\begin{equation}} 
\def\ee{\end{equation}}

\def\kms{\,{\rm {km\, s^{-1}}}} 
\def\msun{{M_\odot}}

\def\HI{\hbox{H~$\scriptstyle\rm I\ $}}

\def\gsim{\lower.5ex\hbox{\gtsima}} 
\def\lsim{\lower.5ex\hbox{\ltsima}} 
\def\gtsima{$\; \buildrel > \over \sim \;$} 
\def\ltsima{$\; \buildrel < \over \sim \;$} 
\def\prosima{$\; \buildrel \propto \over \sim \;$} \def\gsim{\lower.5ex\hbox{\gtsima}} 
\def\lsim{\lower.5ex\hbox{\ltsima}} 
\def\simgt{\lower.5ex\hbox{\gtsima}} 
\def\simlt{\lower.5ex\hbox{\ltsima}} 
\def\simpr{\lower.5ex\hbox{\prosima}}   
  
\def\gtsima{$\; \buildrel > \over \sim \;$} 
\def\ltsima{$\; \buildrel < \over \sim \;$} 
\def\gsim{\lower.5ex\hbox{\gtsima}} 
\def\lsim{\lower.5ex\hbox{\ltsima}} 
\def\simgt{\lower.5ex\hbox{\gtsima}} 
\def\simlt{\lower.5ex\hbox{\ltsima}} 
\def\simpr{\lower.5ex\hbox{\prosima}}

\def\msun{\,{\rm M_\odot}}

\def\E3{{\cal E}_{\rm g}^{III}}

\def\CII{\hbox{C~$\scriptstyle\rm II$}} 
\def\OI{\hbox{O~$\scriptstyle\rm I$}} 
\def\NII{\hbox{N~$\scriptstyle\rm II$}}

\title[FIR lines from high-$z$ galaxies]{FIR line emission from high redshift galaxies}
\author[Vallini et al.]{Livia Vallini$^{1}$\thanks{E-mail:
livia.vallini@sns.it (LV)}, Simona Gallerani$^{1}$, Andrea Ferrara$^{1}$ \& Sunghye Baek$^{1}$\\
$^{1}$Scuola Normale Superiore, Piazza dei Cavalieri 7, 56126, Pisa, Italy}

\begin{document}


\pagerange{\pageref{firstpage}--\pageref{lastpage}} \pubyear{2012}

\maketitle

\label{firstpage}

\begin{abstract}
By combining high resolution, radiative transfer cosmological
simulations of $z \approx 6$ galaxies with a sub-grid
multi-phase model of their interstellar medium we derive the expected
intensity of several far infrared (FIR) emission lines ([\CII] 158
$\mu m$, [\OI] 63 $\mu m$, and [\NII] 122 $\mu m$) for
 different values of the gas metallicity, $Z$. For $Z=Z_\odot$
the $[\CII]$ spectrum is very complex due to the presence of several
emitting clumps of individual size $\simlt3$ kpc; the peak is
displaced from the galaxy center by $\approx 100 \kms$. While the
$[\OI]$ spectrum is also similarly displaced, the [\NII] line comes
predominantly from the central ionized regions of the galaxy. When
integrated over $\sim 500~\rm{km\,\,s^{-1}}$, the $[\CII]$ line flux
is 185~$\rm{mJy~km~s^{-1}}$; 95\% of such flux originates from the
cold ($T\approx 250$ K) \HI phase, and only 5\% from the warm
($T\approx 5000$ K) neutral medium. The $[\OI]$ and $[\NII]$ fluxes
are $\sim$6 and $\sim$90 times lower than the $[\CII]$ one,
respectively. By comparing our results with observations
  of {\it Himiko}, the most extended and luminous Lyman Alpha Emitter
  (LAE) at $z=6.6$, we find that the gas metallicity in this source
  must be sub-solar. We conclude that the $[\CII]$ line from $ z
\approx 6$ galaxies is detectable by the ALMA full array in $1.9 <
t_{\rm ON} < 7.7$ hr observing time, depending on $Z$.
\end{abstract}



\section{Introduction}\label{intro}


High-$z$ galaxies are mainly discovered by means of their Lyman-$\alpha$ emission line \citep[Lyman Alpha Emitters; LAEs, e. g.][]{malhotra2005, shimasaku2006, hu2010, ouchi2010} or through drop-out techniques \citep[Lyman Break Galaxies; LBGs, e.g.][]{steidel1996, castellano2010, bouwens2011, mclure2011}. Both methods are plagued with intrinsic limitations:
the Ly$\alpha$ detection is hampered by the increasingly neutral InterGalactic Medium (IGM), while the source redshift cannot be precisely determined with drop-out techniques; in addition the restframe optical/UV radiation is strongly affected by presence of dust.
It is then important to assess whether other probes, as the far infrared (FIR) metal lines ([\CII],[\OI], [\NII]) originating from the interstellar medium (ISM) of galaxies, could be used to detect new distant sources or better determine the properties of those already discovered. These lines are not affected by \HI or dust attenuation, can deliver the precise redshift of the emitter, and open a window to investigate the structure of the galactic ISM.

Among FIR lines, the $^{2} P_{3/2} \rightarrow$ $^{2}P_{1/2}$ fine-structure transition of ionized carbon [\CII], a major coolant of the ISM, is by far the most widely used to trace the diffuse neutral medium \citep[e.g.][]{dalgarno1972, stacey1991, wolfire1995, lehner2004}. Up to now, high redshift ($z>4$) detections of [\CII] lines have been obtained mainly in sources with high star formation rates (SFRs) \citep[e.g.][]{cox2011, debreuck2011} or in those hosting Active Galactic Nuclei (AGN) \citep[e.g.][]{maiolino2005,gallerani2012}. Recently, \citet{walter2012} put upper limits on the [\CII] luminosity arising from a Gamma Ray Burst (GRB) host galaxy and two LAEs with moderate SFR. Other interesting fine-structure lines are [\OI] 63 $\rm \mu m$, tracing neutral (higher density) gas, and [\NII] 122 $\rm \mu m$ probing the ionized ISM phase. 
[\OI] detections have been reported in two lensed Ultra-Luminous Infrared Galaxies at $z=1.3$ and $z=2.3$ \citep{sturm2010}; $z>4$ nitrogen lines (including the [\NII] 205 $\mu$m) have been detected in quasars and submillimeter galaxies \citep{ferkinhoff2011,nagao2012,decarli2012, combes2012}. The unprecedented sensitivity of ALMA will revolutionize the field allowing the detection of FIR lines from the known ``normal'' population of high-$z$ galaxies \citep[e.g.][and references therein]{carilli2013} as in the case of [\CII] detections in two $z=4.7$ LAEs presented by \citet{carilli2013b}. Therefore, developing models to predict FIR line luminosities and relate them to other physical features such as metallicity, $Z$, and SFR is fundamental to design and
interpret future experiments. 

In this work, we present the first detailed predictions for the intensity of several FIR emission lines ([CII] 158 $\mu m$, [OI] 63 $\mu m$, and [NII] 122 $\mu m$) arising from the ISM in high-$z$ star forming galaxies. Our work is similar in spirit to that of \citet{nagamine2006}, who computed the [\CII] galaxy luminosity function based on a SPH simulation coupled with a sub-grid multi-phase model of the ISM. We improve upon \citet{nagamine2006} work in at least two ways: (a) we concentrate on a single prototypical high-$z$ galaxy, a $z=6.6$ LAE, hence reaching a sufficiently high resolution to properly describe the ISM small-scale density structure; (b) we implement radiative transfer which is crucial to model the intensity of the galactic UV field and the gas ionization structure.


\section{Numerical Simulations}

\begin{figure*}
\centering
\includegraphics[scale=0.3]{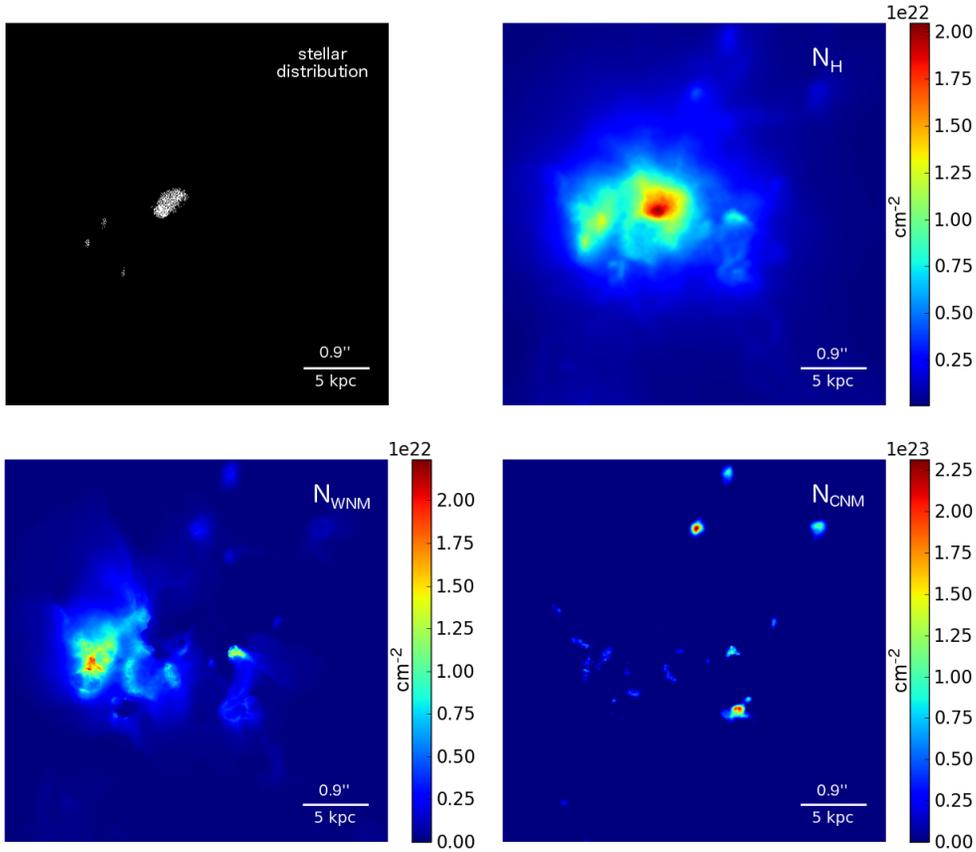}

\caption{\emph{Upper panels}: Projected stellar distribution (left) and hydrogen column density (right). \emph{Lower panels}: warm (left), and cold (right) neutral medium column density. The distribution of WNM is  more diffuse compared to that of CNM which is predominantly found in small ($D\leq 2$ kpc) clumps far from star forming regions.}
\label{columndens}
\end{figure*}

 We run cosmological SPH hydrodynamic simulations using GADGET-2 \citep{springel2005}.
 We use the recent WMAP7+BAO+$H_0$ cosmological parameters: $\Omega_m=0.272$, $\Omega_{\Lambda}=0.728$, $\Omega_b=0.0455$, $h=0.704$, $\sigma_8=0.807$ \citep{komatsu2011}. We simulate a $(10h^{-1}\rm{Mpc})^3$ comoving volume with $2\times512^3$ baryonic+dark matter particles, giving a mass resolution of 1.32 (6.68)$\times10^{5}\msun$ for baryons (dark matter) and gravitational softening $\epsilon = 2 h^{-1}\rm{kpc}$. We select a snapshot at redshift $z=6.6$, and we identify the most massive halo (total mass $M_h=1.17\times10^{11}\msun$, $r_{vir}\approx20$ kpc) by using a Friend-of-Friend algorithm.
We select a $(0.625\,h^{-1}\rm{Mpc})^3$ comoving volume around the center of the halo, and post-processed UV radiative transfer (RT) using LICORICE  \citep{baek2009}. LICORICE uses a Monte Carlo ray-tracing scheme on an adaptive grid. We set the adaptive grid parameter to have a minimum RT size of $ 0.61~h^{-1}\rm{kpc}$. 
Starting from the density field provided by GADGET,  we recompute gas temperature including atomic cooling from the initial temperature $T_0=10^4$ K. The initial ionization fraction is set to $x_{HII}=0$.\\ 
To define the position of the ionizing sources we assume that stars form in those cells characterized by a gas density $\rho \ge \rho_{th}$. We choose $\rho_{th}=1 \,\rm{cm^{-3}}$ in order to reproduce the typical size ($\sim 1-2~\rm kpc$) of star forming regions at $z\approx 6$ \citep{bouwens2004, ouchi2009}, as inferred by UV continuum emitting images.
The projected position of stellar sources is shown in
  white in the upper left panel of Fig. \ref{columndens}. A central large
stellar cluster is clearly visible, along with other 3 minor stellar clumps displaced from the center.
We use the population synthesis code STARBURST99 \citep{leitherer1999}
to obtain the ionizing spectrum of the galaxy. Theoretical works
suggest that high-$z$ galaxies might be relatively enriched ($Z \simgt
0.1~Z_\odot$) galaxies \citep{dayal2009, Salvaterra11}. We adopt $Z =Z_\odot$ as a fiducial value for our study but we also consider a lower metallicity case, i.e. $Z=0.02~Z_\odot$. We assume
a Salpeter initial mass function with a slope of $\alpha=2.35$ in the
mass range 1-100$\msun$, a continuous star formation rate of $10\msun
{\rm yr ^{-1}}$, obtained from the SFR-$M_h$ relation at $z=6.6$
\citep{baek2009,baek2012}. Ionizing UV luminosity is about
$L_{UV}\approx 7 \times 10^{43} \rm{erg\,s^{-1}}$. RT calculations are performed until equilibrium between photoionizations and recombinations is achieved; this occurs within $\approx 10 $ Myr.
The public version of GADGET-2 used in this work does not include the star formation process, neither the radiative cooling, nor supernova feedback. The inclusion of radiative cooling may affect the baryon density profile, enhancing the density towards the center of the galaxy, whereas supernova feedback tends to smooth out density inhomogeneities.  We have checked that the baryon density profile resulting from the simulations used in this work fits well with our previous low resolution simulations which include all these processes \citep{baek2009}. Finally, we note that the large gravitational potential of massive galaxies reduces the effects of SN feedback on star formation, as exemplified by Fig. 1 of \citet{vallini2012} and related discussion.
We interpolate all gas physical properties around the halo center on a fixed $512^3$ grid using the SPH kernel and smoothing length, within a $(0.156\,h^{-1}\rm{Mpc})^3$ comoving volume. We achieve a higher resolution by interpolating on a finer grid as shown in Fig.~6 of \citep{baek2012}. This method also allows us to have continuous density PDF at low and high dense region thus increases the maximum density about 50\% from $64^3$ grid to $512^3$ grid. The resulting hydrogen column density map is shown in the upper right panel of Fig. \ref{columndens}.\\

\begin{figure*}
 \centering
 \includegraphics[scale=0.3]{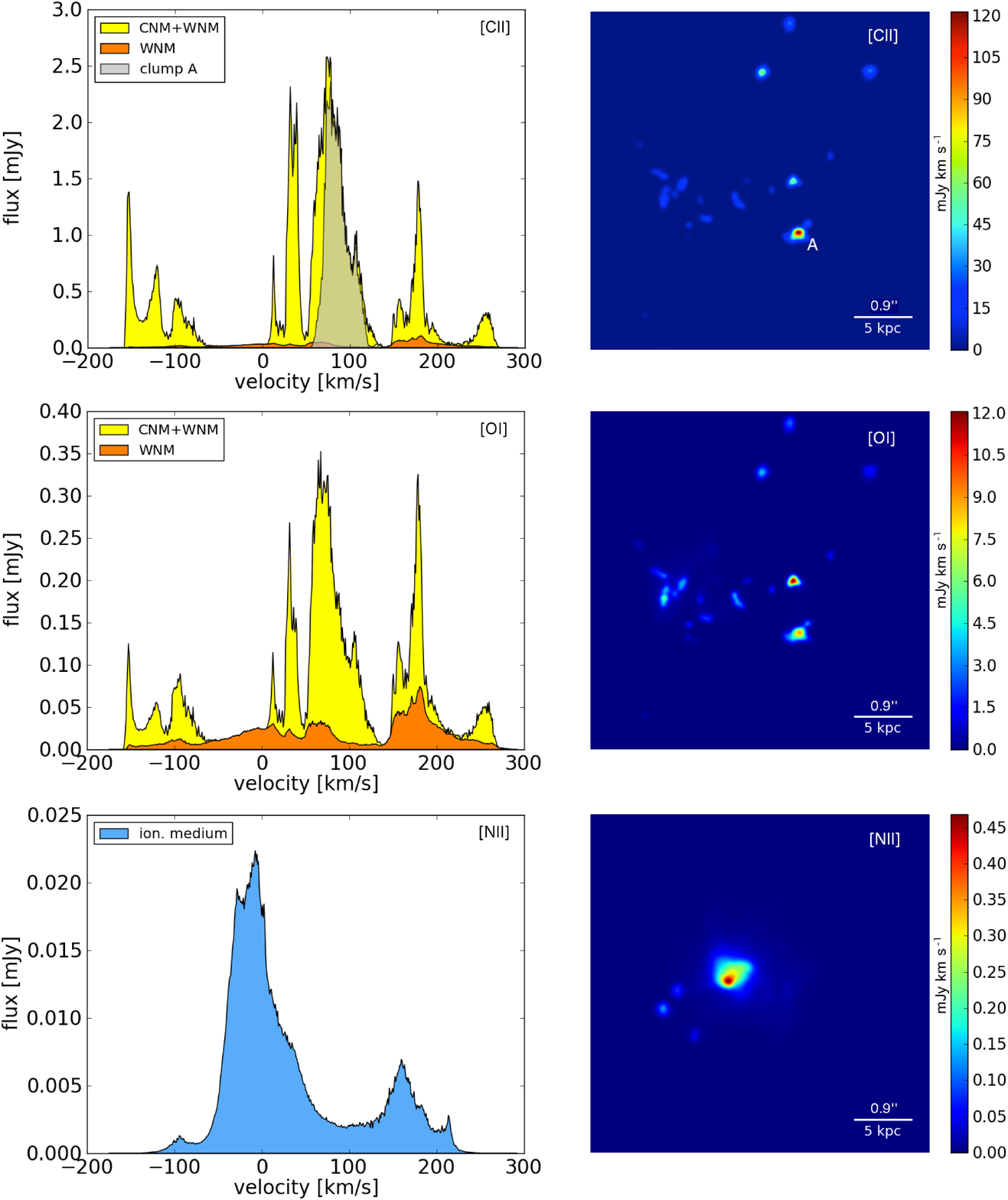}
 \caption{\emph{Left column}: Total (CNM+WNM)  and WNM only (orange) spectrum of $[\CII]$, $[\OI]$ and $[\NII]$ binned in $1.0\,\, \rm km\,s^{-1}$ channels. \emph{Right column}: $[\CII]$, $[\OI]$ and $[\NII]$ maps in mJy km s$^{-1}$ with resolution of $0.1\, \rm{arcsec}$ and integrated over the entire spectral velocity range. The contribution of clump A to the [\CII] spectrum is plotted in gray.}
 \label{results}
\end{figure*}


\section{Multiphase ISM model}


With current computational resources, it is not possibile to self-consistently include sub-kpc scale physics in the above RT simulations. To this aim we adopt a sub-grid scheme based on the model by \citet[][hereafter W95, W03]{wolfire1995, wolfire2003}, in which ISM thermal equilibrium is set by the balance between heating (cosmic rays, X-rays, and photoelectric effect on dust grains) and cooling (H, He, collisional excitation of metal lines, recombination on dust grains) processes (see Table 1 in W95):
\begin{equation}\label{equilibrium}
\mathcal{L}(n, x_e, T)= n^2 \Lambda - n \Gamma = 0,
\end{equation}
where $n \Gamma$ ($n^2 \Lambda$) is the heating (cooling) rate per unit volume $[\rm erg\, s^{-1} cm^{-3}]$, and $n$ is the total gas density. The ISM can be described as a two-phase gas\footnote{Our multi-phase model does not include molecular clouds and therefore emission from dense photodissociation regions (PDRs) which we plan to address in future work.} in which the cold (CNM) and the warm neutral medium (WNM) are in pressure equilibrium. Each cell of the simulated volume is characterized by a pressure $p=(1+x_e) n k_b T$, where $x_e$ is the ionized fraction, taken from the RT simulation output. We determine the density and the temperature of the CNM and WNM by solving Eq. \ref{equilibrium}, coupled with the ionization equilibrium equation. As metal cooling is not implemented neither in GADGET-2, nor in LICORICE the gas cannot cool below $T_{\rm min}\approx7700$ K. We apply the sub-grid ISM multi-phase model only to cells with $T_{min} \le T \le 10^4\,\rm{K}$ where the upper limit is determined by the fact that metals dominate the cooling for temperatures below $10^{4}\,\rm{K}$.
The rates of photoelectric effect and recombination on dust grains strongly depend on the FUV incident radiation. 
The {\bf incident} radiation field ($G$) in the Habing band ($6-13.6 \,{\rm eV}$) is computed at each pixel position $\vec{r}=(x,y,z)$, summing contributions from all sources as following,

\begin{equation}
G(\vec{r})= \Sigma^{n_*}_{i=1}  \frac{\int_{6\,\rm eV}^{13.6\,\rm eV} L_{\nu,i}\rm{d}\nu}{4\pi| \vec{r}-\vec{r_i} |^2},
\end{equation}
where $L_{\nu,i}$ is the monochromatic luminosity per source, $n_*$ is the number of sources, and $\vec{r_i}$ represents their positions. By scaling the flux with the Habing value ($1.6\times10^{-3}\,\rm{erg\,cm^{-2}\,s^{-1}}$) \citep{habing1968} we obtain the standard flux $G_0$.
Within our simulated galaxy we obtain $0.5<\log G_0<5$. 
We find that the mean CNM (density, temperature) is ($\langle n_{\rm{CNM}}\rangle =  50\,\,\rm{cm^{-3}}, \langle T_{\rm{CNM}}\rangle = 250\,\,\rm{K}$, while for the WNM we obtain instead ($\langle n_{\rm{WNM}} \rangle = 1.0\,\,\rm{cm^{-3}}, \langle T_{\rm{WNM}}\rangle = 5000 \,\,\rm{K}$). 
\\

In the lower panels of Fig. \ref{columndens} we show the WNM and CNM column densities. 
The WNM distribution closely traces regions of high ($N_H\approx 10^{22}\,\rm{cm^{-2}}$) total hydrogen column density that are sufficiently far from the central star forming region in order not to become ionized; cold gas lies instead only in small ($D \leq 2\,\rm kpc$) overdense clumps at the periphery of the galaxy. The maps show that cold gas clumps are surrounded by diffuse halos of warm neutral medium.


\subsection{FIR emission lines}


For each simulated cell we estimate the line luminosities $L_i=\epsilon_i V_{cell}$, where the emissivity, $\epsilon_i$, is given by:
\begin{equation}\label{cii}
\epsilon_{i}(n,T)= \Lambda^{H}_{i} \chi_{i} n^2 + \Lambda^{e-}_{i} \chi_{i} x_e n^2,
\end{equation}
where $n$ and $T$ are the density and temperature of the WNM/CNM, $\Lambda^{H}_{i}$ ($\Lambda^{e-}_{i}$) is the specific cooling rate due to collision with H atoms (free electrons) taken from \citet{dalgarno1972}, and $\chi_{i}$ is the abundance of the i-th species.
The $[\NII]$ line traces the ionized medium since its ionization potential ($14.5\,\rm{eV}$) exceeds 1 Ryd. Therefore, it provides a complementary view of the ISM with respect to the $[\CII]$ and $[\OI]$ lines. The $[\NII]$ cooling rate due to collisions with free electrons is:
\begin{equation}
\epsilon_{\rm N_{II}}(n,T)=\frac{A h \nu}{n_{c}} \frac{g_u/g_l}{1 + \left[(g_u/g_l) + 1 \right] (n_e/n_{c})} \chi_{\rm N_{II}} x_e n^2,
\end{equation}
where $A=7.5 \times 10^{-6} \rm{s^{-1}}$ is the Einstein coefficient, $\nu$ is the frequency for the $^{3}P_2 \rightarrow$ $^{3}P_1$ transition, $h$ is the Planck constant, $g_u/g_l$ is the ratio of the statistical weights in the upper and lower levels, and $n_{c} = 300 \,\rm{cm^{-3}}$ is the $[\NII]$ critical density for $T=10^4\,\rm{K}$. We finally compute the observed flux by integrating along 
the line-of-sight also accounting for the gas peculiar velocity field obtained from the simulation.


\section{Results}


In Fig. \ref{results} we show the predicted $[\CII]$ 158~$\mu m$, $[\OI]$ 63~$\mu m$ and $[\NII]$ 122~$\mu m$ emission for the spectral resolution of our simulations ($1.0\,\,\rm{km\,s^{-1}}$), a beam resolution of $0.1$ arcsec and $Z=Z_{\odot}$, along with the maps obtained by integrating the spectra over the full velocity range $-200 < v < 300\,\,\rm{km\,s^{-1}}$.

The $[\CII]$ spectrum contains considerable structure due to the presence of several emitting CNM clumps distributed over the entire galaxy's body ($\sim20$ kpc). The individual sizes of the clumps are however much smaller ($\simlt 3\,\,\rm kpc)$. The peak of the spectrum reaches $\sim2.5\,\,\rm{mJy}$ and it is displaced from the center of the galaxy by about 100 $\rm{km\,s^{-1}}$. This is due to the fact that the gas within the central kpc of our galaxy is highly ionized by the massive stars that form there. We find that 95\% of the total [\CII] flux originates from the CNM, and only 5\% from the WNM. For the $[\CII]$ emission line we obain a flux of 185~$\rm{mJy~km~s^{-1}}$, integrating over $\sim 500~\rm{km\,\,s^{-1}}$. 

In Fig. \ref{results} we plot in grey the spectrum extracted by integrating over a circular area of $\sim 2~\rm kpc$ radius, centered on the component labeled A in the map. It dominates the peak of the $[\CII]$ spectrum (30\% contribution to the total emission), with the remaining $\sim70\%$ coming from less luminous substructures. This is an important point as with high spatial resolution observations a substantial fraction of the $[\CII]$ emission may remain undetected.
The FWHM of the main peak is $\sim50$ km s$^{-1}$, consistent with the marginal detection of [\CII] in high-$z$ LAEs \citep{carilli2013}.
We have computed FIR line intensities also for a metallicity $Z=0.02~Z_{\odot}$. In this case, the $[\CII]$ and $[\OI]$ intensities drop by a factor of $\sim 1000$ and $\sim 300$, respectively, whereas the $[\NII]$ flux is reduced by a factor of 50.
While the WNM emission is $\propto Z$, at very low $Z$ CNM is practically absent, since the lower metal content makes the CNM phase thermodynamically unfavorable. A thorough analysis of the relative fraction of the emission arising from CNM and WNM as a function of $Z$ will be adressed in a forthcoming paper.

The $[\OI]$ spectrum has a shape similar to that of $[\CII]$ since for both emission lines we are taking into account the emission arising from the neutral phase of the ISM. In the case of [\OI], 75\% of the total flux arises from the CNM and 25\% from the WNM.
The maximum value of the $[\OI]$ flux is $\sim0.35\,\,\rm{mJy}$.
The $[\NII]$ emission line reaches a maximum flux of $0.022\,\,\rm{mJy}$ at $v=0$.
This line traces the ionized phase of the ISM, and the bulk of its emission arises from the center of the galaxy where the ionizing field intensity is higher. In conclusion, the $[\OI]$ and $[\NII]$ fluxes are $\sim$6 and $\sim$90 times lower than the $[\CII]$ one.


\section{Comparison with observations}


\subsection{LAE observations}
As pointed out in the introduction, FIR line observations in high-$z$ sources have been carried out mainly in quasars and sub-millimeter galaxies. Recently, \citet{walter2012} have tried to detect the [CII] emission in {\it Himiko}, one of the the most luminous LAEs at $z=6.6$ \citep[][]{ouchi2009}. However, they end up only with a $1\sigma$ upper limit of $0.7\,\,\rm{mJy\,km\,s^{-1}}$.

\begin{figure}
\centering
\includegraphics[scale=0.3]{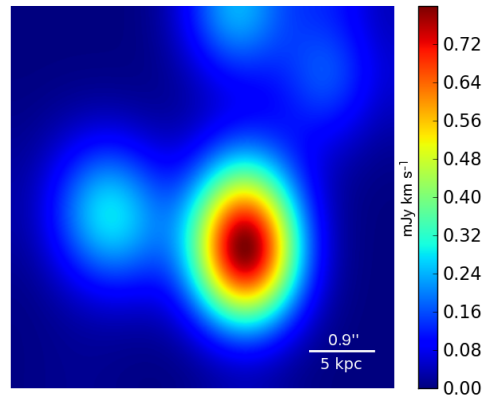}
\caption{Synthetic map of $[\CII]$ emission in mJy km s$^{-1}$ integrated over a velocity channel of width $ = 200\,\,\rm{km~s^{-1}}$, and smoothed to an angular resolution of $2.27'' \times 1.73''$ to allow comparison 
with \citet{walter2012} observations.}
\label{comp_walter}
\end{figure}

The large size of the {\it Himiko} Ly$\alpha$ emitting nebula ($\geq 17~\rm kpc$) makes this object one the most massive galaxies discovered at such high redshifts \citep{ouchi2009, wagg2012}. From this point of view, {\it Himiko}'s properties closely resemble those of the prototypical galaxy selected from our simulation. Moreover, the radius of the region within which we distributed the stars ($\sim 1-2$~kpc) is consistent with the {\it Himiko} half-light radius (1.6 kpc) observed by Ouchi et al. (2009). Other properties of {\it Himiko} are poorly constrained. The SFR is highly uncertain and its value strongly depends on the diagnostics used to infer it: SED fitting gives $\simgt34\,\rm{M_{\odot}\,yr^{-1}}$, UV luminosities yields $=25^{+24}_{-12}\,\rm{M_{\odot}\,yr^{-1}}$; the Ly$\alpha$ line implies $36\pm 2 \,\rm{M_{\odot}\,yr^{-1}}$.  As for the metallicity, \citet[][]{ouchi2009} suggest $Z=[1-0.02]~Z_{\odot}$ as a plausible range, i.e. consistent with the one we have chosen for our analysis.
For a fair comparison with the Plateau de Bure Interferometer data by \citet{walter2012}, we smooth our $[\CII]$ simulations to a beam resolution of $2.27'' \times 1.73''$, and we produce channel maps of $200\,\,\rm{km\,s^{-1}}$ width. 
In Fig. \ref{comp_walter} we show the map with the largest signal achieved. 
We find that, for $Z=Z_\odot$ the maximum intensity is $\sim0.72\,\rm{mJy\,km\,s^{-1}}$, slightly exceeding the observed upper limit by \citet{walter2012}; thus, we can put a solid upper limit on \textit{Himiko}'s metallicity  $Z < Z_{\odot}$. This shows the potential of FIR lines in obtaining reliable metallicity measures in high-$z$ galaxies.

\subsection{Low redshift observations}

Haro 11 (H11), a nearby ($z\sim 0.02$) dwarf galaxy \citep{cormier2012}, is considered a suitable local high-$z$ galaxy analog. Through PACS observations of the $[\CII]$, $[\OI]$ and $[\NII]$ lines, \citet{cormier2012} measured a size of $\sim3.9\,\,\rm{kpc}$ for the H11 star forming region, a value which is comparable to the size of the clump A shown in the upper-most right panel in Fig.\ref{results}. These authors also estimate the relative contribution to the observed FIR lines from the diffuse (neutral/ionized) medium and PDRs. They found that $\sim80\%$ of the $[\CII]$ and $[\NII]$ emissions come from the diffuse medium, while the $[\OI]$ mostly originates from PDRs. We scale the luminosities of the predicted FIR emission lines to the H11 luminosity distance ($D_L\sim 88$) and metal abundances \citep{cormier2012}. 
For a fair comparison with the data, taken from Tab. 2 of \citep{cormier2012}, we compute [\CII], [\OI], and [\NII] spectra by integrating over a region of $\sim12$ kpc in diameter, which corresponds to an angular size of $30''$ at the H11 redshift. 
For $[\CII]$ and $[\NII]$ lines our model predicts a flux  corresponding to $20\%$ of the observed one. 
For what concerns $[\OI]$, we recover only $3\%$ of the observed flux. However, we recall that the contribution of PDRs, not included in our model, might be non-negligible.


\section{ALMA predictions}


In Table \ref{alma}, we plot the expected total fluxes for the FIR emission lines considered, varying the metallicity between $Z_{\odot}$ and $0.02~Z_{\odot}$. In the solar metallicity case a $[\CII]$ $\sim 5\sigma$ detection over four 25~$\rm km~s^{-1}$ channels requires a sensitivity of 0.2 mJy, which translates into an observing time of $t_{\rm ON}=1.9$ h with the ALMA full array. We note that the predicted fluxes are sensitive to the actual value of $Z$,  implying that a $[\CII]$ line detection can strongly constrain LAE metallicities. On the other hand, this implies that LAEs characterized by metallicities $Z<0.5~Z_{\odot}$ would require a  long observing time ($t_{\rm ON}>7.7$ h) to be detected even with the ALMA full array.    

\begin{table}
\centering
    \begin{tabular}{rccc}
\toprule
\hline
& \multicolumn{3}{c}{Integrated flux [mJy km s$^{-1}$]} \\
&$Z=Z_{\odot}$ & & $Z=0.02\,Z_{\odot}$\\ 
\midrule
 $[\CII]$&185& &0.2\\ 
 $[\OI]$&30& &0.1\\ 
 $[\NII]$&2& &0.04\\
\bottomrule
    \end{tabular}
\caption{Integrated flux over 500 km s$^{-1}$ channel, arising from our simulated source for $Z=Z_{\odot}$ and $Z=0.02\,Z_{\odot}$.}
\label{alma}
\end{table}


\section{Summary and conclusions}


We have presented the first attempt to predict the intensity of several FIR emission lines ($[\CII]$ 158 $\mu m$, $[\OI]$ 63 $\mu m$, and $[\NII]$ 122 $\mu m$) arising from the ISM of high-$z$ star forming galaxies. We combined RT simulations of a $z=6.6$ galaxy with a sub-grid multi-phase model to predict the density and temperature of the cold and warm neutral phase of the diffuse ISM. 
We find that warm neutral medium lies in overdense regions located sufficiently far from the central star forming clump where the strong ionizing UV field does not allow the presence of neutral gas.  Cold gas resides instead in more dense clumps. The physical properties of the cold and warm neutral medium deduced here are in agreement with previous studies \citep[e.g.][]{wolfire1995, wolfire2003}: the mean density (temperature) of the CNM (WNM) gas are $\langle n_{\rm{CNM}}\rangle =  50\,\,\rm{cm^{-3}}, \langle T_{\rm{CNM}}\rangle = 250\,\,\rm{K}    $, and $\langle n_{\rm{WNM}} \rangle = 1.0\,\,\rm{cm^{-3}}, \langle T_{\rm{WNM}}\rangle = 5000 \,\,\rm{K}$, respectively.

Assuming $Z=Z_{\odot}$, our model predicts for the $[\CII]$ emission line a flux of 185~$\rm{mJy~km~s^{-1}}$, integrating over $\sim 500~\rm{km\,\,s^{-1}}$. The $[\OI]$ and $[\NII]$ fluxes are $\sim$6 and $\sim$90 times lower than the $[\CII]$ one, respectively. We have investigated also the case of $Z=0.02~Z_{\odot}$. At this metallicity, the $[\CII]$ and $[\OI]$ intensities drop by a factor of $\sim 1000$ and $\sim 300$, respectively, while the $[\NII]$ flux is reduced by a factor of 50. 

In the case of $Z=Z_{\odot}$, we have found that 95\% (75\%) of the $[\CII]$ ($[\OI]$) emission arises from the cold neutral medium (CNM) of the ISM, and the remaining 5\% (25\%) from the warm neutral phase. In the lower metallicity case, the fluxes of the $[\CII]$ and $[\OI]$ emission lines drop abruptly since the lower metal content does not allow the presence of CNM phase. 
As a caveat we note that the [\OI] 63 m$\mu$ line could be optically thick \citep[e.g.][]{vasta2010}.
The intensity of the $[\NII]$ line, instead, scales linearly with the metallicity, since it arises from the ionized medium.

Interestingly, the $[\CII]$ and $[\OI]$ lines are shifted with respect to the $[\NII]$ line, as a consequence of the fact that they originate from different regions: while the ionized medium, which is traced by the $[\NII]$ line, is located close to the center of the galaxy, the neutral gas, from which the $[\CII]$ and $[\OI]$ lines originate, is predominantly located at large galactocentric radii. This result can explain the shift between the $[\CII]$ and $[\NII]$ lines observed in some high-$z$ galaxies \citep[e.g.][]{nagao2012}.
We have compared our predictions with observations of FIR emission lines in high-$z$ and local star forming galaxies. At $Z=Z_{\odot}$, our model slightly exceeds the $1\sigma=0.7\,\,\rm{mJy\,km\,s^{-1}}$ upper limit on the $[\CII]$ intensity found in \textit{Himiko} through PdBI observations \citep{walter2012}. This result suggests that the gas metallicity in this source must be sub-solar. Our results are also marginally consistent with $[\CII]$, $[\OI]$, and $[\NII]$ observations of Haro 11 \citep{cormier2012}, a suitable high-$z$ galaxy analog in the Local Universe. In this case, our model predicts a flux which is $\sim$20\% ($\sim$3\%) of the observed one in the case of $[\CII]$ and $[\NII]$ ($[\OI]$) emissions. 

We underestimate the observed flux in Haro11 as a non-negligible fraction of their flux may be provided by dense PDRs not included yet in our study. In particular the $[\OI]$ line is expected to originate primarily from PDRs \citep{cormier2012}. We defer the inclusion of PDRs in a forthcoming paper. 

According to our findings, the $[\CII]$ emission line is detectable with the ALMA full array in $1.9 < t_{\rm ON} < 7.7$~hr in star forming, high-$z$ galaxies with  $Z_{\odot}>Z>0.5~Z_{\odot}$. We emphasize again that our predictions provide a solid lower limit to the expected FIR emission lines flux. 
  
Finally, the results presented in this work might be very useful to FIR line intensity mapping studies. In fact, our model represents a valid tool to calibrate the intensity of these lines depending on the different properties of the first galaxies, such as the metallicity and the SFR. Since the mass of the CNM increases in weaker FUV radiation field environments, is it is likely that the specific emission from FIR emission lines as the $[\CII]$ and $[\OI]$ could increase towards fainter galaxies. We leave a dedicated study of this effect to future work.

\section*{Acknowledgments} 
We thank F. Combes, D. Cormier, S. Madden, and  T. Nagao for useful discussions and comments.  

\label{lastpage}

\bibliographystyle{mn2e}
\bibliography{bibliography}

\end{document}